\documentclass{article}

\usepackage{chicago}
\newcommand{\ie}{\mbox{i.\,e.\,\ }}
\newcommand{\iec}{\mbox{i.\,e.\,}}

\newcommand{\egc}{\mbox{e.\,g.\,}}

\newcommand{\dr}[1]{\ensuremath{\mathrm{d} #1\,}}
\newcommand{\mc}[1]{\ensuremath{\mathcal{#1}}}

\newcommand{\vbv}[2]{\ensuremath{\frac{\delta #1}{\delta #2}}}

\newcommand{\real}{\ensuremath{\mathrm{Re}}\,}

\newcommand{\ket}[1]{\ensuremath{\left|  #1 \right\rangle}}
\newcommand{\bra}[1]{\ensuremath{\left\langle #1 \right|}}
\newcommand{\bk}[2]{\ensuremath{\left\langle #1 | #2 \right\rangle}}

\newcommand{\tpk}[2]{\ensuremath{\ket{#1}\!\otimes\!\ket{#2}}}

\newcommand{\matel}[3]{\ensuremath{\bra{#1} #2 \ket{#3}}}
 
\newcommand{\op}[1]{\ensuremath{\widehat{\textsf{\ensuremath{#1}}}}}
\newcommand{\opad}[1]{\ensuremath{\op{#1}^{\dagger}}}
\newcommand{\id}{\op{\mathsf{1}}}
\newcommand{\denop}{\ensuremath{\rho}}

\newcommand{\tr}{\textsf{Tr}}
\newcommand{\nrm}{\frac{1} {\sqrt{2} } }

\newcommand{\be}{\begin{equation}}
\newcommand{\ee}{\end{equation}}
\newcommand{\e}[1]{\mathrm{e}^{#1}}

\renewcommand{\Pr}{\mathrm{Pr}}

\begin{document}

\title{Spontaneous Symmetry Breaking in Finite Quantum Systems: a decoherent-histories approach}
\author{David Wallace\thanks{School of Philosophy, University of Southern California; email \texttt{dmwallac@usc.edu}}}
\maketitle
\begin{abstract}
Spontaneous symmetry breaking (SSB) in quantum systems, such as ferromagnets, is normally described as (or as arising from) degeneracy of the ground state; however, it is well established that this degeneracy only occurs in spatially infinite systems, and even better established that ferromagnets are not spatially infinite. I review this well-known paradox, and consider a popular solution where the symmetry is explicitly broken by some external field which goes to zero in the infinite-volume limit; although this is formally satisfactory, I argue that it must be rejected as a physical explanation of SSB since it fails to reproduce some important features of the phenomenology. Motivated by considerations from the analogous classical system, I argue that SSB in finite systems should be understood in terms of the approximate decoupling of the system's state space into dynamically-isolated sectors, related by a symmetry transformation; I use the formalism of decoherent histories to make this more precise and to quantify the effect, showing that it is more than sufficient to explain SSB in realistic systems and that it goes over in a smooth and natural way to the infinite limit.
\end{abstract}

\section{A story about symmetry breaking}\label{story}

There is a standard account of spontaneous symmetry breaking (SSB) in quantum theory that is physically intuitive, explanatorily powerful, and predictively effective, and it goes like this. To begin with, consider a single nonrelativistic particle of mass $m$, moving in one dimension in a smooth potential $V(x)$. In full generality, very little can be said about this particle's dynamics. But if we now suppose that the potential has a single minimum at $x=x_0$, we can expand the potential around that minimum:
\be \label{basic-potential}
V(x_0+\eta)=V(x_0) + \frac{1}{2}V''(x_0) \eta^2 + O(\eta^3).
\ee
And if the potential is sufficiently slowly changing, then at least for the lowest-lying energy eigenstates, we can neglect the terms in $(x-x_0)^n$ for $n>2$ and approximate the system as a harmonic oscillator, with energy levels
\be
E_n = V(x_0) + \hbar (n+1/2)\omega.
\ee
where $\omega^2=V''(x_0)/m$.
The well-known methods of perturbation theory now allow us to make successive corrections to these energy levels due to the higher-order terms in the expansion of $V(x_0)$ around $x=x_0$, and also to determine when the harmonic oscillation starts to break down entirely.

Now suppose the potential is symmetric under $x\rightarrow -x$. If it has a single minimum at $x=0$, the analysis above applies unchanged, with the extra detail that the perturbative corrections are also symmetric under this transformation, and so the structure of the spectrum of energy eigenstates shows that symmetry: in particular, since the spectrum is nondegenerate, each eigenstate is individually symmetric or antisymmetric with respect to the transformation. (And so in particular, the position-space wavefunction $\psi$ of an eigenstate satisfies $\psi(-x)=\pm \psi(x)$.)

If instead there is a minimum at $x_0>0$, there will be another minimum at $x=-x_0$. For definiteness, we can consider the potential
\be \label{doublewell}
V(x) =  \frac{\lambda}{4!}x^4 -\frac{\mu^2}{2}x^2 + \frac{3\mu^4}{2 \lambda}
\ee
with minima at $x=\pm x_0\equiv\pm(6\mu^2/\lambda)^{1/2}$ (and where the constant is just for convenience, to enforce $V(x)=0$ at the minima). Expanding around, say, the minimum at $x=+x_0$, we get
\be
V(x_0+\eta)=\frac{1}{2}(2\mu)^{1/2} \eta^2 + O(\eta^3)
\ee
which --- if the minima are sufficiently widely separated, and the function's slope is sufficiently gentle --- we can again treat as a perturbed harmonic oscillator. So at least for low-lying energy eigenstates, the system can (apparently) be treated as degenerate, with two ground states localised around $x=\pm x_0$ and two towers of oscillations, one built on each ground state. There will be perturbative corrections due to the higher-order terms in the expansion, and non-perturbative corrections due to quantum tunnelling between the minima, but (again, apparently) these have only minor effects on the physics provided the system is not too excited. However, the expansion of the potential around a given minimum will not be symmetric under $\eta\rightarrow -\eta$. So long as we confine our attention to one set of low-lying states, the symmetry will not be manifest. It will instead be detectable only in certain relations between the higher-order coefficients in the expansion.

In particular, at sufficiently low temperatures the statistical mechanics of the system will display a certain asymmetry despite the symmetry in $V(x)$: the statistical-mechanical state of the system will be \emph{either} a canonical-ensemble state built out of the excitations localised around the left-hand potential well, \emph{or} the equivalent state built out of the excitations localised around the right-hand well. As the temperature is increased, this will break down, and eventually (and, for this small system, in a rather fuzzy and imprecise way) the energy eigenstates making a significant contribution to the canonical ensemble will be spread across both wells.

Qualitatively similar phenomena occur when we move from a single quantum particle to a quantum field theory\footnote{For details on quantum field theory as it is used in high-energy physics, see (e.g.) \citeN{peskinschroeder}, Weinberg~\citeyear{weinberg,weinbergqft2}, \citeN{duncanQFT}; for quantum field theory as it is used in solid-state physics, see \citeN{altlandsimons}; for a conceptual introduction aimed at non-specialists, see \citeN{wallaceqfthandbook}.} --- say, to a scalar theory defined (either through canonical quantization, or directly via the path integral) by Lagrangian
\be \label{phi-fourth}
\mc{L}=\int \dr{x}^3 \left( \frac{1}{2}\partial_\mu\phi(x) \partial^\mu\phi(x) - V(\phi(x))\right)
\ee
where $V$ is again an arbitrary smooth potential and where the field theory is defined in some \emph{finite} region, say of volume $L^3$ (and where we use units with $c=1$). If $V$ has a single minimum at $\phi=\phi_0$, we can expand around that minimum ---
\be
V(\phi_0+\eta) = V(\phi_0)  + \frac{1}{2}V''(\phi_0)\eta^2 + O(\eta^3)
\ee
 --- and treat the field theory, to first approximation, as a free field theory with  Lagrangian
 \be
 \mc{L}_0 = \int \dr{x}^3 \left( \frac{1}{2}\dot{\eta}(x)^2 - \frac{1}{2}(\nabla \eta(x))^2 - m^2 \eta(x)^2\right).
 \ee
  The ground state $\ket{\Omega}$ of that theory --- the \emph{vacuum}, in particle-physics terminology --- can be treated --- again, to first approximation --- as the ground state of a collection of uncoupled oscillators corresponding to the modes of the free field; its excitations, one family per mode, are characterised by a \emph{dispersion function} $\omega(k)$, so that $\hbar \omega(k)$ is the energy of an excitation of momentum $\hbar k$. In our particular example, $\hbar \omega(k)=(m^2+\hbar^2k^2)^{1/2}$; in more general theories (in particular, in non-Lorentz-covariant field theories like those encountered in solid-state physics) the dispersion function can have a more general form. In any case, the excitations can be interpreted (at least approximately) as \emph{particles}, and the non-quadratic corrections to $V(\phi_0+\eta)$ can be treated perturbatively as interactions between the particles, giving rise to scattering. The expectation value of the field operator on the vacuum is \be
 \matel{\Omega}{\op{\phi}(x)}{\Omega}= \phi_0.
\ee
  (This whole framework requires some kind of short-distance cutoff to be mathematically well-behaved, and the exact issue of the relation between field and particle has its own subtleties, but these issues are orthogonal to those I want to consider here; see Wallace~\citeyear{wallacecritique,wallaceqfthandbook} and references therein for further discussion.) 
  
  If the potential function has a minimum at $\phi=0$ and is symmetric under $\phi\rightarrow -\phi$, that symmetry will be manifest in the particle structure of the theory and in particular in the multiparticle interactions. If the potential is symmetric but has minima away from $\phi=0$ --- for definiteness again, if it has the quartic  form (\ref{doublewell}) --- we expect \emph{two} vacuum states $\ket{\Omega+}, \ket{\Omega-}$, with field expectation values 
  \be
   \matel{\Omega\pm}{\op{\phi}(x)}{\Omega\pm}= \pm \phi_0. 
  \ee
 The excitations around each vacuum can be described in particle language again, but the symmetry is now hidden, and cannot be seen explicitly in multiparticle interactions. 
 
 If we consider the statistical mechanics of this theory, at low energies we expect again two different thermal equilibria, each consisting in the free-particle approximation of a gas of excitations around a given vacuum. As the temperature increases the relevant excitations will begin to spread out of the separate potential wells, so that eventually the symmetry will be restored. Depending on the system it is at least plausible that this will constitute a \emph{phase transition}, characterised by changes in the functions of state or their derivatives which become sharper and sharper as $L$ gets larger, tending smoothly in the limit as $L\rightarrow \infty$ to discontinous change.
 
 This somewhat heuristic and perturbative description of SSB in field theory can be put on a firmer footing using the \emph{effective action} formalism (\citeNP{colemanweinberg,weinberg1973,dolanjackiw}). Its details will not be salient for this paper (for a pedagogical introduction in the zero-temperature regime, see \citeN[ch.11]{peskinschroeder}), but it allows us to define a temperature-dependent functional --- the ``effective action'' ---
 \be
 \Gamma[\phi;T]
 \ee
 of a classical field $\phi$ with the properties that:
 \begin{enumerate}
 \item The effective action has the same symmetries as the underlying field theory.
\item If $\phi_c$ solves
\be
\left.\vbv{\Gamma[\phi;T]}{\phi}\right|_{\phi=\phi_c}=0
\ee
then $\phi_c(x)$ is the expectation value of $\op{\phi}(x)$ in the thermal-equilibrium state at temperature $T$. In particular, at $T=0$, $\phi_c(x)$ is the vacuum expectation value of $\op{\phi}(x)$.
\item
The eigenfunctions of the operator
\be\label{second-derivative}
\left.\frac{\delta^2 \Gamma[\phi;T]}{\delta\phi(x)\delta\phi(y)}\right|_{\phi=\phi_c}
\ee
(understood as an operator on real functions on $\Re^4$)
determine the elementary excitations of the field theory at temperature $T$, and in particular the mass of the particles.
 \end{enumerate}
 So the thermal-equilibrium state of the field theory breaks symmetry at temperature $T$ iff $\Gamma[\phi;T]$ is extremised for nonzero $\phi$ (and in this circumstance there are multiple extrema, and so multiple thermal equilibria). In particular the ground state is degenerate if $\Gamma[\phi;0]$ has multiple extrema. If symmetry is not spontaneously broken then the symmetries of the effective action are also symmetries of $(\ref{second-derivative})$ and hence of the particle spectrum. If it is spontaneously broken, the particle spectrum does not display the symmetry. And the signature of a phase transition is a theory where the extrema of the effective action are nonzero below some critical temperature $T_c$ and zero above that temperature. 
 
 Finally, if the field theory has a (global) symmetry that is continuous, spontaneous breaking of that symmetry leads to \emph{Goldstone bosons}: elementary excitations whose dispersion relations satisfy $ \omega(0)=0.$ (In relativistic theories this corresponds to massless particles.) One example can be given just by taking our scalar field to be complex-valued, with Lagrangian
 \be
 \mc{L}=\int \dr{x}^3 \left( \frac{1}{2}\partial_\mu\phi^*(x) \partial^\mu\phi(x) - V(|\phi(x)|)\right)
 \ee
so that it is invariant under $\phi\rightarrow \e{i\theta}\phi$. For another illustrative example\footnote{See \citeN[ch.2]{andersonbasicnotions} and references therein; or, for a more elementary introduction, \citeN[ch.23]{ashcroftmermin}.} consider nonrelativistic physics with finite particle density. At low temperatures, the equilibrium state will usually be some kind of crystal, and hence spontaneously breaks translation symmetry; the associated Goldstone bosons are the \emph{phonons}, quanta of vibration, which have a dispersion relation satisfying $\lim_{k=0}\omega(k)=0.$ The statement that Goldstone bosons are a universal consequence of spontaneous breaking of a (global) continuous symmetry is \emph{Goldstone's theorem}; it can be proved via the effective action by consideration of (\ref{second-derivative}) and through several other ways\footnote{See, \egc,\citeN[section 19.2]{weinbergqft2}, \citeN[p.1136-1140]{wittenqft}, and references therein.}, and has as its central premise the existence of a vacuum state in which symmetry is broken.
 
 To repeat: this account is physically intuitive, explanatorily powerful, and predictively effective. So it is a little disturbing to note that it cannot, strictly speaking, be correct.
 
 \section{Non-degeneracy of the ground state}\label{groundstate}

Consider again the example of a single particle in a double-well potential, and for the moment imagine modifying that potential to introduce an infinitely-high potential barrier at $x=0$. (Formally, we can replace $V(x)$ with
\be
V'(x) = V(x) + K \delta(x)
\ee
where $\delta$ is the Dirac delta function,
and take the limit as $K\rightarrow \infty$.)
In this case, the dynamics entirely decouples into left-side and right-side dynamics. On the left, we will have an infinite tower of energy eigenstates $\psi_{L,n}$ ---
\be
\op{H}\psi_{L,n}(x) =E_n \psi_{L,n}(x)
\ee
 --- with $\psi_{L,n}(x)=0$ for $x>0$.
To first approximation these eigenstates will be harmonic-oscillator eigenstates with energy $E_n=\hbar (n+1/2) (V''(-x_0)/m)$. The non-quadratic part of the potential will perturb them away from this simple form and make $O(\hbar^2)$ corrections to the energy; in principle these perturbations will include the effect of the infinite barrier but in practice the details of this barrier will make very little difference for small $n$ since $\psi_{L,n}(x)$ will be constrained to be very small for $x\simeq 0$ in any case.

On the right, there will be a similar family of eigenstates $\psi_{R,n}$. Given the symmetry, we have $\psi_{R,n}(x)=\psi_{L,n}(-x)$, and in particular $\psi_{R,n}(x)=0$ for $x<0$; the symmetry also entails that the eigenvalue of $\psi_{R,n}$ is also $E_n$. So we have exact degeneracy not only of the ground state but of all excitations: energy eigenstates appear in pairs.

Now take out the infinite potential barrier, so that the $\psi_{L,n}$ states are confined on the left side only by the finite potential barrier between the two wells. It is an elementary result of quantum mechanics that no finite potential can \emph{exactly} confine a wavefunction. It may take a  long time, but eventually any such wavefunction will tunnel through to the right-hand side. Which is to say that $\psi_{L,n}$ is not an eigenstate of energy --- hence, not a stationary state --- once the barrier is removed. 

To explore this further: if we take symmetric or antisymmetric combinations of the $\psi_{L,n}$ and $\psi_{R,n}$, we expect those states to be very close to invariant under tunnelling, even for arbitrarily long times. So those symmetric and antisymmetric combinations are --- to a very good approximation --- the eigenstates of energy, but there will be an energy gap between them, generated by tunnelling effects and completely lifting the ground-state degeneracy. Or, to put it another way, we expect that (for sufficiently low-lying energies) the exact eigenstates of energy come in symmetric and antisymmetric pairs $\psi_{\pm,n}$ satisfying $\psi_{\pm,n}(x)=\pm \psi_{\pm,n}(-x)$, such that the combinations $\psi_{+,n}\pm \psi_{-,n}$ are very close to our original states $\psi_{L,n}$, $\psi_{R,n}$.

This heuristic argument can be established by direct calculation of the energy gap, using instanton methods. See \citeN[ch.7]{coleman} for the details, but in outline: the low-lying energy states, and in particularly the ground state, do come in pairs $\psi_{\pm,n}(x)$, with
energies
\be
E_{\pm,n}= (n+1/2) \hbar \omega + O(\hbar^2) \pm \frac{\kappa \hbar \omega}{2} (S/\hbar)^{1/2}\e{-S/\hbar}.
\ee
Here, the first term is the ordinary harmonic-oscillator energy. The $O(\hbar^2)$ term represents perturbative corrections to the harmonic oscillator which are the same for $E_{+,n}$ and $E_{-,n}$. And the exponential term gives the energy-level splitting. $S$ itself is given by
\be \label{instanton}
S = \int_{-x_0}^{+x_0}\dr{x}\sqrt{2mV(x)}
\ee
and can be interpreted as the action of the \emph{instanton} connecting the two classical minima $\pm x_0$. ($\kappa$ is a dimensionless constant of order unity, whose exact form will not be relevant to us.)
The splitting term is in general extremely small --- and, in particular, it tends to zero as $\hbar \rightarrow 0$ faster than any power of $\hbar$, and so faster than any perturbative correction --- but it is not zero, and that is enough to break the degeneracy. 

Essentially the same effects occur in quantum field theory with a double-well potential. Again, given an infinite potential barrier between the two wells, we expect exact degeneracy, and two families of excitations --- interpretable as particles --- built one upon each ground state. But again, if the barrier is only finitely high then tunnelling can occur between the two potential wells and the degeneracy is lifted. The energy gap between (low-lying) pairs of states is given (via a straightforward generalisation of the method of instantons) by
\be\label{energygap-field}
\Delta E = \frac{\kappa m}{2}(S/\hbar)^{1/2}\e{-S/\hbar}
\ee
where now
\be\label{instanton-field}
S=L^3 \int_{-\phi_0}^{+\phi_0}\dr{\phi} \sqrt{2V(\phi)}.
\ee 
The case of (apparently) spontaneously broken continuous symmetry is qualitatively similar but differs in the details. No energy barrier confines the system to any particular value of the average field strength, so any localised state will spread out in the same way that a wave-packet state of a free particle spreads out. In the case of the complex scalar field in particular, if we write $\phi=\rho\exp(i\theta)$ then the Lagrangian becomes
\be
\mc{L}= \int \dr{x}^3\left(\frac{1}{2} \rho(x)^2 \partial_\mu \theta(x) \partial^\mu \theta(x) +\frac{1}{2} \partial_\mu \rho(x) \partial^\mu \rho(x) - V(\rho(x))\right).
\ee
In the approximation where interactions can be neglected, we can decouple the $\theta$ and $\rho$ modes and get 
\be
\mc{L}_\theta = \frac{1}{2}\int \dr{x}^3 \rho_0^2 \partial_\mu \theta(x) \partial^\mu \theta(x)
\ee
where $\rho_0$ is the value of $\rho$ that minimises $V(\rho)$. (This is one form of the \emph{non-linear sigma model}; see, \egc, \citeN[ch.19]{weinbergqft2}, \citeN[ch.IV-1]{dynamicsofstandardmodel} and references therein.) If we define 
\be
\bar \theta = \frac{1}{L^3}\int \dr{x^3} \theta(x)
\ee
and write $\theta(x)=\bar\theta + \delta \theta(x)$, then the Lagrangian decouples further:
\be\label{nonlinear-sigma}
\mc{L}_\theta = \frac{\rho_0^2L^3}{2} \dot{\bar \theta}^2 + \frac{1}{2}\int \dr{x}^3 \rho_0^2 \partial_\mu \delta\theta(x) \partial^\mu \delta \theta(x).
\ee
The average value of the phase just has the Lagrangian for a free particle with mass $(\rho_0^2 L^3)^{-1}$, moving on the unit circle; clearly, when quantized this system has a nondegenerate ground state with wavefunction $\Psi[\bar\theta]=(2\pi)^{-1/2}$. Similarly, given a finite-size crystal, the wavefunction of the center of mass will spread out freely until it is uniformly spread over the region accessible to the crystal, so again the ground state, which is time-independent, must be nondegenerate. (For a similar analysis of the harmonic crystal, see \citeN{wezelbrink}.)

So: in all the systems we considered in section \ref{story}, the lowest-energy state is nondegenerate. (And, although the arguments I have given involve various approximation schemes, this can be proven rigorously for one-particle systems and at least a wide class of field theories; see \citeN[pp.1125-1130]{wittenqft} for details.) And this seems to make Section \ref{story}'s analysis, which relied on ground state degeneracy, completely untenable.

\section{The return of degeneracy in infinite systems}

On a purely formal level, degeneracy can in a sense be recovered by considering \emph{infinite} quantum systems, using the tools of algebraic quantum field theory, or AQFT.\footnote{For reviews of AQFT, see, \egc, \citeN{haag}, \citeN[ch.1-5]{ruetschebook}, or \citeN{halvorsonqftencyclopedia}; for conceptual discussions of SSB from an AQFT viewpoint, see Ruetsche~(\citeyearNP{ruetschejohnny}, \citeyearNP{ruetschebook} ch.12-14) and \citeN{bakerhalvorsonssb}.)} In AQFT, a quantum field theory is characterised in the first instance by the algebraic structure of the field observables, without stipulating in advance a representation of those field observables as Hilbert-space operators. Particular models of the quantum field theory then correspond to particular representations of those observables as operators on some Hilbert space. This algebraic approach is in principle available for ordinary quantum-mechanical systems, but in practice adds little since (under weak assumptions; cf \citeN[ch.3]{ruetschebook}) any two such representations are unitarily equivalent, \iec are related by some unitary transformation. In infinite-volume quantum field theories\footnote{And also in finite-volume quantum field theories without the short-distance cutoff I introduced in section \ref{story}, though it is doubtful whether this has any physical significance; see Wallace~\citeyear{wallacecritique,wallaceqfthandbook} for further discussion of this point.} this is not the case: there are many (indeed, uncountably many) unitarily inequivalent representations. In particular, we can understand SSB in this framework as the existence of inequivalent representations, differing by the spatially-averaged value of the field whose expectation value breaks the symmetry. In each such representation the ground state remains nondegenerate, but the symmetry transformation is represented not by a unitary transformation \emph{of} each representation, but by a non-unitary map \emph{between} representations. 

But of course, real condensed-matter systems are manifestly \emph{not} infinite; \emph{perhaps} particle-physics systems should be thought of as infinite, but they are at any rate not fields in infinite \emph{Minkowski} spacetime, as they are represented in AQFT. In both cases, the spatially-infinite description seems something that we should regard as a useful idealisation or approximation of some finite system (or some finite patch of a larger cosmological system) and not something to be taken literally.\footnote{Feynman: ``[I]t is not often that experiments are done under the stars. Rather they are done in a room. Althought it is physically reasonable that the walls have no effect, it is true that the original problem is set up as an idealisation. It is no more satisfactory idealization to remove the walls to infinity than to replace them by perfect mirrors far away. The mathematical rigor is wasted in the first idealization, since the walls are not at infinity.'' \cite[pp.93-4]{feynmanhibbs}.} From this perspective (urged upon us by \citeN{Earman2004} and \citeN{butterfieldemergence2}, physical phenomena that occur in the infinite limit ought to be taken seriously only when they are in some sense approximated by physical phenomena that occur in finite systems, with the approximation getting better as the size of the finite system increases. We see this in phase transitions: their strict mathematical characterisation is in terms of a discontinuity in their state functions or the derivatives of those functions; this discontinuity will not occur in any finite system; but the function will change increasingly rapidly and will tend towards a discontinuous function as the system volume tends towards infinitity. (See \citeN{butterfieldemergence2} for further discussion of this example.)

It is not at all clear how to understand SSB in this way. The ground state of our systems remains resolutely non-degenerate, however big the system is, and so is invariant under the symmetry operation. So the expectation value of the supposedly symmetry-breaking field, with respect to the ground state, must remain exactly zero no matter how big the system is, and does not in any obvious sense tend towards the nonzero value that is obtained in an infinite system and that is calculated perturbatively, and empirically successfully, using effective-action methods. Nor can the effective action even really be defined for infinite systems: it is proportional to the spatial volume of the system and so is infinite for infinite systems.

To summarise: a naive consideration of symmetry breaking in finite systems leads to the prediction of a multiplicity of ground states transforming nontrivially under the symmetry, a degenerate collection of excitations built on that ground state, and associated phenomenology (such as symmetry restoration at high temperatures, and Goldstone bosons). Those predictions are confirmed empirically, to great accuracy, and yet the naive considerations they rest on are \emph{false}, and the ground state of finite systems shows no degeneracy. Degeneracy does occur in genuinely infinite systems, and those genuinely infinite systems reproduce at least some of the phenomenology of SSB. But the temptation to regard those infinite systems as the correct description of the physics (a position discussed sympathetically by Ruetsche~\citeyear{Ruetsche2003,ruetschebook} and Batterman~\citeyear{batterman,Batterman2005}) is at least prima facie in conflict with the obvious fact that solid-state systems are of finite size, and the (somewhat less obvious) fact that particle-physics systems are only accurately described as fields on Minkowski spacetime when confined to finite regions; it is also in conflict with empirically-very-powerful calculational tools like the effective action, which apparently require finite volume to be well defined.

In the remainder of the paper I will explore how SSB can, after all, be understood in finite systems. But in this exploration it will be important to remember that the infinite limit cannot be simply disregarded. Infinite-system models do succeed in describing SSB, and it is implausible that this is a coincidence: our task is not just to understand the process in finite systems but also to understand in what sense those systems' behaviour tends towards the symmetry-breaking behaviour of the infinite systems in the limit of large size.

\section{A partial solution: symmetry breaking from the environment}

Suppose we return to the double-well potential (\ref{basic-potential}) and add a linear symmetry-breaking term:
\be
V(x) \rightarrow V_J(x) = V - Jx
\ee
(with, for definiteness, $J>0$). The effect will be to explicitly break the symmetry: to a first approximation, the classical ground state at $x=+x_0$ will decrease in energy by $Jx_0$ and the ground state at $x=-x_0$ will increase by a similar amount. (The linear term will also slightly displace the minimum of the potential.) Provided that $Jx_0$ is large compared to the tunnelling-induced energy gap $\kappa \hbar \omega (S/\hbar)^{1/2}\e{-S/\hbar}$ (as defined by (\ref{instanton})) between the pairs of energy eigenstates, this term will drastically change the shape of the energy eigenstates: now the ground state will be very close to a harmonic-oscillator ground state at $x\simeq +x_0$, and there will be an excited state very close to a harmonic-oscillator ground state at $x \simeq -x_0$. Indeed, if  $J x_0 \ll \hbar \omega$, the excitation energy required to add a quantum of energy to either potential well, then this will be the \emph{first} excited state. And, for at least the lower-lying excited states, the same will occur: the spectrum will continue to consist of closely-spaced pairs of eigenvalues, but the eigenstates of the eigenvalue pair around $(n+1/2)\hbar \omega$ will be very close to the $n$th excited states of harmonic oscillator potentials localised at $\pm x_0$.

So the introduction of even this very small asymmetry restores the picture of the double-well potential as describing a system which can be localised in either one of two harmonic-oscillator potentials. The only difference between this picture and the picture of section \ref{story} is that there is not exact degeneracy between the two oscillators: the family of excitations of the left-hand oscillator is higher in energy than that of the right-hand oscillator by $\simeq 2 J x_0$. (As usual, for sufficiently excited states this account will begin to break down as the states cease to be localised even approximately in one well or the other.)

Essentially the same story plays out for the continuum version of the double well potential, though the effect is more dramatic: we modify the Lagrangian (\ref{phi-fourth}) to
\be\label{perturbed-lagrangian}
\mc{L}_J = \int \dr{x}^3 \left( \frac{1}{2}\partial_\mu\phi(x) \partial^\mu\phi(x) - [V(\phi(x))- J \phi(x)]\right) 
\ee
which will  localise the ground state (in field-space, not physical space) around $\phi=+\phi_0$ provided that $J \phi_0$ is large compared to the energy gap given by (\ref{energygap-field}). Since this energy gap decreases exponentially with system volume, as that volume gets larger the perturbing field can become smaller and smaller while still localising the ground state.

In for the case of continuous symmetry breaking, the appropriate modification is to add to the potential a term $-J \mathrm{Re}(\phi)$ (that is: proportional to the real part of $\phi)$. The part of the Lagrangian controlling the average phase is then modified to
\be
\mc{L}_{\bar\theta} = \frac{\rho_0^2L^3}{2} \dot{\bar \theta}^2 + J \rho_0 L^3 \cos \bar\theta
\ee
and for small $\bar\theta$ this is just a harmonic-oscillator potential:
\be
\mc{L}_{\bar\theta} \simeq \frac{\rho_0^2L^3}{2} \dot{\bar \theta}^2 - \frac{J \rho_0L^3}{2} \bar\theta^2 + \mbox{constant}.
\ee
Elementary quantum mechanics then tells us that for $|\bar \theta| \ll \pi$, the ground state wavefunction is
\be
\Psi_\Omega(\bar\theta) \propto \exp ( - (J \rho_0)^{1/2}L^3 \bar \theta^2/2\hbar)
\ee
which clearly becomes increasingly narrow as $L$ increases, tending to a delta-function as $L\rightarrow \infty$.

The symmetry-breaking term also gives a way of making sense of effective-action methods. The effective-action formalism is in fact slightly more general than I described in items (1) to (3) in section \ref{story}: solving the equation
\be
\left.\vbv{\Gamma[\phi;T]}{\phi}\right|_{\phi=\phi_c}=J
\ee
gives the vacuum expectation value of the field given external field $J$ as described by the Lagrangian (\ref{perturbed-lagrangian}), and the second variational derivative of $\mc{L}$ evaluated at this solution gives the particle spectrum given that external field. (So the results of section \ref{story} are the special case where $J=0$.) For sufficiently large $J$, the expectation value can be nonzero, \iec the symmetry is explicitly broken. If we define an effective action per unit volume,
\be
\tilde \Gamma = \frac{1}{L^3}\Gamma,
\ee
then we can take $L$ and $J$ to zero simultaneously while keeping $J \phi_0 \gg \e{-SL^3/\hbar}$, and get a well-defined effective action in the joint limit of exact symmetry and infinite volume. It is at least plausible  to suppose (though, in the absence of physically-realistic interacting models of the AQFT axioms, not rigorously proven) that this effective action describes the physics of the actually-infinite, AQFT models discussed in the previous section.

These mathematical results suggest a physical way of understanding symmetry breaking in finite systems. If we suppose that in realistic systems the system's environment is never exactly uniform, it becomes possible to suppose that for any specific system, the symmetry is explicitly broken by some environmental asymmetry. And since for sufficiently large systems the required asymmetry can be very small indeed, the supposition that such asymmetry exist might seem innocent --- indeed, the contrary assumption of exact environmental asymmetry becomes extremely implausible.

This external-perturbation approach to symmetry breaking has been widely advocated both in textbooks and  pedagogical articles (\egc, \citeNP{wezelbrink}) and in conceptual and foundational work (\egc, \citeNP{landsman2013}, \citeNP{jfraserssb}; see also further references in Fraser, \emph{ibid}). But I do not think it is ultimately viable.

One potential problem is that in \emph{high-energy-physics contexts} it is not so obvious that it makes sense to assume any ``external'' field in this sense. If we suppose that symmetry breaking occurs dynamically in the early Universe, normal physics practice is to treat that symmetry breaking as a \emph{source} of subsequently observed inhomogeneities in the Universe, not to simply posit those inhomogeneities outright. And (unlike in condensed-matter contexts) we cannot straightforwardly regard $J$ as representing some unanalysed external perturbation, because there is nothing obviously external to do the perturbing.

However, I do not think this is fatal to the external-perturbation approach, for a technical reason: there do not appear to be any spontaneously-broken exact symmetries in high-energy physics, at least in the sense considered in this article. (The most prominent application of global SSB in high-energy physics is the identification of pions as pseudo-Goldstone bosons associated to a spontaneous breaking of chiral symmetry in QCD --- cf \citeN[ch.19.4-19.5]{weinbergqft2}, \citeN{schererchiral}, \citeN[ch.VII]{dynamicsofstandardmodel} and references therein --- but there the symmetry is \emph{explicitly} broken by the masses of the quarks, and that explicit symmetry-breaking term will certainly suffice to induce a unique noninvariant ground state.) Indeed, it is commonly said in high-energy physics (cf, \egc, \citeN[pp.489-492]{duncanQFT}) that there \emph{are} no exact global symmetries: apparent global symmetries are ``accidental'' symmetries that arise from the fact that symmetry-breaking terms would have to be nonrenormalisable and so are too small to be readily detected at energies accessible to us. High-energy physics does involve spontaneous breaking of exact \emph{local} symmetries,\footnote{Specifically, through the Higgs mechanism for electroweak symmetry breaking: see, \egc, \citeN[ch.21.3]{weinbergqft2}} and cosmology seems to involve spontaneous breaking of translation and rotation symmetry in \emph{non}-equilibrium contexts\footnote{For technical details see, \egc, \citeN{weinbergcosmology} and references therein; for some consideration from a philosophical point of view, see \citeN{wallacecosmology}.} but neither phenomenon lies within the scope of those discussed here.

A much more serious problem for the external-perturbation approach comes from the more mundane context of condensed-matter physics. Consider, for definiteness, a ferromagnet like an iron bar: above its critical temperature, the spins of the iron atoms are randomly allocated, but when it is cooled below the critical temperature then nearby spins line up: that is, rotational symmetry is spontaneously broken. The external-perturbation approach, taken completely literally, assumes the existence of some small constant $J$ (a magnetic field, in effect) that explicitly breaks the symmetry, so that all of the spins in the iron bar line up in the direction of $J$. But this is not what happens when iron is actually cooled (unless experimenters explicitly choose to introduce a significant magnetic field). In practice, the iron bar forms ``domains'', typically $\sim 10^4-10^6$ atoms across; atoms within a given domain are aligned with one another, but the various domains are independently and randomly aligned.

Of course, it is fairly straightforward to imagine how to modify the external-perturbation theory to account for this. We could suppose that (i) the external perturbation varies randomly from point to point, and (ii) it is constant over scales of order the domain size, so that within each domain the external perturbation determines the spin direction. This is now a quite substantial hypothesis about ambient random magnetic fields; it also implies that cooling ferromagnets are extraordinarily sensitive detectors of those random magnetic fields. Were we to take the hypothesis seriously, we should be allowing ferromagnets to cool, and measuring their domain sizes, in various different conditions where no macroscopically-detectable magnetic field is present, to see if we can detect overall anisotropies, and variations in fluctuation scale, in this mysterious background field.

But in fact, domain size does not vary in this way: it is an intrinsic feature of ferromagnets, and indeed one that is pretty well understood theoretically. Domain theory (see, \egc, \citeN{kittelgalt} and references therein) explicitly derives the size of ferromagnetic domains as a consequence of the trade-off between the energetic costs of domain wall formation and of dipole-dipole interaction between domains. Absent a miraculous coincidence whereby space is filled with a tiny randomly-fluctuating magnetic field whose correlation length just so happens to accurately match the exact correlation length independently determined by domain theory, the account of SSB and domain formation in the ferromagnet seems to explicitly require the \emph{absence} of any magnetic field strong enough to explicitly break symmetry and overcome the internal features of the ferromagnet that force domains to be randomly aligned on an intrinsically-determined scale.

So, while the introduction of an external field provides a satisfactory \emph{mathematical} account of the large-volume limit of systems with SSB, it does not seem possible to interpret it as the \emph{physical} mechanism, at least not in some important physical cases.

\section{Symmetry breaking as a correlation phenomenon: the classical case}\label{classicalcase}

To see what I think is the right way to understand SSB in finite systems, it will help to think about the probability distribution of properties of those systems --- and, for the moment, to ignore quantum mechanics and treat those systems as classical. So consider a classical model of a ferromagnet --- where the system is a finite lattice of spins, each of which can point (for simplicity) either up or down --- and suppose that we measure the value of \emph{one} of those spins. 

If the system's state does not spontaneously break symmetry, then of course we expect the spin to be up with probability 50\%, down with probability 50\%. What if it \emph{does} break symmetry? Then \emph{either} all the spins (or all the spins in some domain, at any rate) are up, \emph{or} all of them are down. But given that the overall dynamics is symmetry preserving, and given \emph{ex hypothesi} that no external perturbation has explicitly broken the symmetry, the probability of each symmetry-breaking configuration should be equal.\footnote{This is admittedly a little handwaving; I will argue for it more carefully shortly.} So \emph{whether or not} the symmetry is broken, we predict the same statistics for each spin. And the argument generalises: given any property of the system, we expect the probability distribution over values of that property to be invariant under the symmetry, whether or not symmetry is spontaneously broken.

So where does the symmetry breaking show up? In these places:
\begin{enumerate}
\item In correlations between spins at different spatial points. In unbroken symmetry, the spin of one particle carries no information about the spin of other particles: spin values at different points in the lattice are uncorrelated. In spontaneously broken symmetry (and in the idealisation where domains are infinite), the spin of one particle completely determines --- \iec, is perfectly correlated with --- the spin of all the other particles.
\item In the symmetry-breaking properties of each individually-likely value of some collective properties, as a result of the above. The spatially-averaged spin of a large region of the ferromagnet, if the symmetry is unbroken, will with probability $\sim 1$ be very close to zero. The spatially-averaged spin of that same region, given spontaneously broken symmetry, will with probability $\sim 1$ be very close either to $+1$ or to $-1$. (Note that whether or not the symmetry is spontaneously broken, the \emph{expected} value of the spatially-averaged spin is exactly zero.)
\item In correlations between spins of the same point at different times. In unbroken symmetry, we expect the spin of any given point to be frequently re-randomised by interactions with its neighbors, so that spins at time $t$ and time $t'$ are uncorrelated provided that $(t-t')$ exceeds some rerandomisation timescale that is normally very short compared to other dynamically-relevant timescales. In spontaneously broken symmetry, we expect the spins at $t$ and $t'$ to be perfectly correlated even for very large values of $(t-t')$. (And, as a corollary, we expect the values of collective properties, like spatially-averaged spins, to be likewise almost-perfectly correlated over long times.)
\end{enumerate}

It is straightforward to see explicitly how this is realised in classical statistical mechanics. Consider, to begin with, the simple case of one particle in a one-dimensional potential well described by a potential $V$ symmetric under $x \rightarrow -x$. If that particle is in thermal contact with a reservoir, statistical mechanics describes it by the canonical distribution: the probability, at any given time, of the particle having position $q$ and momentum $p$ 
is
\be
\Pr(q,p) = \frac{1}{Z(T)}\e{-p^2/2mk_B T}\e{-V(q)/k_B T}.
\ee
This distribution is clearly symmetric in the spatial variable $q$: the expected value of particle position is zero, whether $V$ describes a single-well potential with minimum at $0$ or a double-well potential with minima at $\pm x_0$.

However, if in the latter case the height $V(0)-V(x_0)$ of the potential barrier between the two minima is large compared to $k_B T$, then the multi-time correlations in the two cases will be radically different. Because if so, in the double-well case a particle localised on the left-hand side has probability $\sim 1$ of having a momentum too small to cross the potential barrier and end up on the right hand side at any later time. So if we define the quantity $X(t)$, which takes value $+1$ if the particle is on the right-hand side of the system at time $t$ (\iec, if $q(t)>0$) and $-1$ if it is on the left-hand side of the system ($q(t)<0$), we predict that $\langle X(t)\rangle=0$ in both cases, but that $\langle X(t)X(t')\rangle$ quickly tends to zero in the single-well case, but is equal to $\sim 1$ even for large $(t'-t)$ in the double-well case.

As the temperature increases, of course, this will cease to be the case. If the potential barrier energy is small compared to $k_B T$ then with high probability, the particle will have sufficient energy to pass over the potential barrier with little delay, and so the correlation will fall off rapidly with time whether the potential has a single or a double minimum.

In effect, we can think of the canonical distribution of the system, at sufficiently low temperatures, as being written as
\be
\Pr(q,p) = 0.5 \times \Pr_L(q,p) + 0.5 \times \Pr_R(q,p)
\ee
where $\Pr_L$ and $\Pr_R$ are, respectively, the appropriate canonical ensembles for a particle trapped on the left or the right of an infinite potential barrier at $x=0$. Because there is no such barrier, these separate distributions are not \emph{quite} time-invariant, but to a very good approximation we can consider the system as being in one of two thermal states, with probability 0.5 of each. Using the language of measurement for a moment, if we observe this system then our first observation will (among other things) determine which of the two thermal states the system is in; all subsequent measurements will be relative to that state. But all of this is captured by the single, symmetric, canonical ensemble of the whole system, and need not be put in by hand.

The situation is conceptually still clearer, if technically more complicated, if we consider a classical version of the field-theory case (\ref{phi-fourth}) with a double-well potential. (Since this theory needs to be discretised at high momenta in order to have a well-defined statistical mechanics, it is closely related to the classical ferromagnet we considered above.\footnote{For a more precise description of the relation between them see, \egc, Binney~et al~\citeyear[chs.6-7]{binney}.}) At absolute zero, the canonical distribution for this system is a sum of delta functions: the system has state $\phi(x)=+\phi_0$ with probability 0.5 and $\phi(x)=-\phi_0$ with probability 0.5. So the expected value of $\phi(x)$ for any $x$ is zero, but there is perfect correlation between $\phi(x,t)$ and $\phi(x',t')$ for any $x,x',t,t'$. At temperatures above zero, but still sufficiently small, the canonical distribution of the system will assign probability $\sim 0.5$ to the system's state being a small fluctuation around $\phi(x)=+\phi_0$, and probability $\sim 0.5$ to it being a similar fluctuation around $\phi(x)=-\phi_0$. And the probability (again, as determined by the canonical distribution) of a given state transitioning from the vicinity of one minimum to the vicinity of another in any reasonable timescale will be negligible. Again, we can think of the canonical ensemble of the system as an equally-weighted sum of two canonical ensembles, each \emph{very nearly} time-invariant and each describing a system localised around one or other minimum. In any observational context, the first observation serves to determine which minimum is realised; given the negligible probability of the system transitioning between minima in any reasonable timescale, subsequent observations are determined by the thermal state relevant to the actual minimum.

In this macroscopic system (as opposed to the purely microscopic case of the single particle) we can also use the language of Boltzmannian statistical mechanics (cf, \egc, \citeN{goldsteinboltzmann}, \citeN{lebowitz07}) to describe what is going on. In that language, we partition the phase space of a system (more precisely, each energy hypersurface of that phase space) into `macrostates', each characterised by approximately-constant values of collective physical quantities like (in this case) spatially-averaged field strength over some macroscopic region. On the normal Boltzmannian account of equilibration, for any given energy one such macrostate (the ``equilibrium macrostate'') will have a vastly larger volume than the rest, and so in the absence of other conserved quantities it is reasonable to expect that the system will make its way into that macrostate and remain there for a very long time. The validity of this assumption about the equilibrium macrostate can be checked in the canonical formalism by looking at the variance of the spatially averaged quantities according to the canonical distribution. If it is sufficiently small, the probability of these macroquantities being close to their most-probable value is very large, and --- since the canonical distribution is uniform on each energy hypersurface --- this at least loosely speaking corresponds to the equilibrium macrostate as being dominant in volume.

For this very reason, the straightforward Boltzmannian hypothesis --- \iec, that there is a unique equilibrium macrostate dominating the phase space --- will fail for this system. What we can expect instead (for sufficiently low energies) is that (i) the overwhelming majority of the volume of the hypersurface is split between two equal-volume macrostates, one corresponding to each minimum; (ii) states initially localised in one of these macrostates almost all (as measured by phase space volume) avoid crossing over to the other macrostate unless we wait an extremely long time. In this case, the Boltzmannian account of equilibration will predict that the system will end up in \emph{one or other} of the two equilibrium macrostates. If the probability distribution over non-equilibrium initial conditions is itself invariant under the symmetry, each equilibrium has a 50\% chance of being realised.

To summarise: at least in the classical case, there is a perfectly clear account of SSB to be given that is compatible both with finite systems, and with probability distributions that are required to be invariant under the symmetry.

\section{Correlation probabilities for a quantum particle}\label{singleparticle}

In the light of these observations, let us revisit the ground state of the \emph{quantum} systems we first considered in Section \ref{story}. In the first instance, consider the one-dimensional particle in a symmetric potential well. On symmetry grounds it is clear that the position of the particle is equally likely to be measured at $x>0$ or $x<0$, irrespective of whether we have a single-well or double-well potential. But investigating the multi-time correlation functions for the system is more difficult, since quantum states cannot simply be treated as probability distributions. For the moment, let us adopt an operational approach and suppose that we measure, at successive times $0,\tau,2\tau, \ldots$, whether the particle is on the left or right side of the system. 

If we begin with the single-well harmonic oscillator, we can avoid technical complexities by confining our attention to the two lowest-lying energy eigenstates:
\be
\psi_0(x) = \frac{1}{\pi^{1/4}}\e{-m \omega x^2/2}
\ee
and
\be
\psi_1(x) = \frac{(2 m \omega)^{1/2}}{\pi^{1/4}}x\e{-m \omega x^2/2}.
\ee
The sum and difference of these functions,
\be
\psi_R(x)=\frac{1}{\sqrt{2}}(\psi_0(x)+\psi_1(x))\,\,\,\,\,\psi_L(x)=\frac{1}{\sqrt{2}}(\psi_0(x)-\psi_1(x))
\ee
are fairly well-localised on the right and left sides of $x=0$ respectively: if the system's state is $\psi_R$, the probability of it being found with $x>0$ is $(1/2 + (2\pi)^{-1/2}) \simeq 0.899$. So to a reasonable first approximation we can replace measurement of whether the particle is on the left or right with measurement in the $\psi_L,\psi_R$ basis.

Next, let us model a measurement in this basis by the following unitary interaction (adopting Dirac notation):
\be
\tpk{\psi_L}{\mbox{---}}\rightarrow \tpk{\psi_L}{L}\,\,\,\,\,\tpk{\psi_R}{\mbox{---}}\rightarrow \tpk{\psi_R}{R}.
\ee
Here, $\ket{\mbox{---}}, \ket{L}$ and $\ket{R}$ represent, respectively, states of some external measurement device that is in the ``ready'' state, that has recorded the particle to be in state $\psi_L$, and that has recorded the particle to be in state $\psi_R$, and the overall interaction is assumed to be so quick (relative to the dynamical timescales of the oscillator itself, \iec $\omega^{-1}$) that we can idealize it as instantaneous. Introducing two such measurement devices and allowing the system to evolve unitarily between the first measurement at $t=0$  and the second at $t=\tau$, we find the state of system+devices after that second measurement to be
\begin{eqnarray}
\nonumber\label{two-measurements}
\ket{\psi_2} &=& \nrm\cos(\omega \tau)(\ket{\psi_L}\otimes \tpk{L}{L} + \ket{\psi_R}\otimes \tpk{R}{R})
\\ &+& \nrm\sin(\omega \tau)(\ket{\psi_L}\otimes \tpk{R}{L}+\ket{\psi_R}\otimes \tpk{L}{R}).
\end{eqnarray}
To extend to three measurements, we can define a condensed notation where $\ket{XYZ}\equiv \ket{\psi_Z}\otimes \ket{X}\otimes \ket{Y}\otimes \ket{Z}$. Then the quantum state after three measurements is
\begin{eqnarray}
\nonumber \label{three-measurements}
\ket{\psi_3} & = & \nrm\cos^2(\omega \tau)(\ket{LLL}+\ket{RRR}) + \nrm \sin^2(\omega \tau)(\ket{LRL}+\ket{RLR})\\
&+& \frac{1}{2}\cos(\omega \tau)\sin(\omega \tau)(\ket{LLR}+\ket{RRL}+\ket{LRR}+\ket{RLL}).
\end{eqnarray}
Comparison of these two expressions reveals that the ``probabilities'' that they define do not behave like normal probabilities. Consider the following protocols: (I) measure at time 0, make no measurement at time $\tau$, measure at time $2\tau$. (II) measure at each of times $0,\tau,2\tau$. The probability of obtaining (say) the result ``left, left'' in protocol (I) is (applying the Born rule to (\ref{two-measurements}), with $\tau \rightarrow 2\tau$)
\be
\Pr(L\mbox{?}L)=\frac{1}{2}\cos^2(2\omega \tau).
\ee
The probabilities of obtaining ``left,right,left'' and ``left,left,left'' in protocol (II) (applying the Born rule to (\ref{three-measurements})) are 
\[
\Pr(LLL)=\frac{1}{2}\cos^4(\omega \tau)= \frac{1}{8}(1+\cos(2\omega \tau))^2
\]
\be
\Pr(LRL)=\frac{1}{2}\sin^4(\omega \tau)= \frac{1}{8}(1-\cos(2\omega \tau))^2
\ee
so that 
\be
\Pr(LLL)+\Pr(LRL) = \frac{1}{4}(1 + \cos^2(2\omega \tau))
\ee
which in general does not equal $\Pr(L?L)$.

This, of course, is nothing but quantum interference: there is interference between the $LLL$ and $LRL$ histories such that the sum of their individual probabilities is not necessarily the probability of the more coarse-grained history. When interference is present, it is not mathematically possible to interpret the probabilities defined by our operational protocol as the probabilities that the unmeasured system actually has the measured values of the quantities being measured. 

If we apply a similar analysis to the double-well potential, \emph{formally} we get the same answer. We can define $\psi_L$ and $\psi_R$ in terms of the ground state and the first excited state (as we have already seen, these states are very well approximated, for sufficiently widely separated wells, by the ground states of simple harmonic oscillators centred at the minima of the two wells); the formulae just given will again apply to correlation measurements; the only difference is that instead of $\omega$, we use the frequency gap
\be
\omega_I=\kappa \omega (S/\hbar)^{1/2}\e{-S/\hbar}
\ee
that (as we discussed in Section \ref{groundstate}) can be calculated by instanton methods.
However, there is a crucial \emph{substantive} difference. If the wells are sufficiently separated that $S/\hbar\gg 1$, then $\omega_I \ll \omega$. It is then possible to choose $\tau$ such that $\omega^{-1} \sim \tau \ll \omega_I^{-1}$: that is, the time between measurements is comparable to the oscillation timescales of the individual wells but much less than $1/\omega_I$, the tunnelling timescale for transition between wells. In this circumstance, the probability of obtaining different results on different measurements becomes negligibly small: if the system is measured and found to be on the left hand side, it will continue to be found on the left hand side in any future measurements. As such there is no appreciable interference, and thus no formal barrier --- at least as far as these measurements are concerned --- to interpreting the ground state of the double-well oscillator as a probabilistic mixture of ``L'' and ``R''.

This is suggestive: just as in the classical case, there seems to be a difference between single-well and double-well potentials that manifests in high correlation between particle locations at different times. Indeed, the difference between the two cases seems to manifest not only in different values of the correlations between outcomes, but in the formal validity or invalidity of attributions of probabilities to sequences of outcomes in the first place. In effect, in the single-well case, if we want to study the system on the timescales on which its dynamics is non-trivial, we have to treat its quantum state as a description of the actual, and highly non-classical, physical state. In the double-well case, provided we are not interested in timescales long enough for tunnelling to matter, we are at least formally justified in treating the state as a probabilistic mixture of two single-well-potential quantum states, just as in the classical case. 

In the rest of the paper I will argue that these observations can be developed into a fully satisfactory understanding of SSB in finite systems.

\section{Decoherence and consistent histories: a brief review}

The questions of whether, when, why, and to what degree the quantum state can be treated as a probability distribution is closely tied to the quantum measurement problem: indeed (at least in my view; see \citeN{wallaceorthodoxy} for more details) the most accurate statement of the measurement problem is exactly this set of questions. As such, our consideration of SSB has now taken us into deep and contentious waters.\footnote{The idea that there might be some link between the measurement problem, inequivalent representations, and SSB is not new; it has been previously explored by, \egc, \citeN{hepp}, \citeN{saundersaqft}, \citeN{landsman1991}, and Breuer \emph{et al}~\citeyear{breueretal}. However, these papers develop a quite distinct line of thought from the one I develop here: in particular, appeal to the infinite limit appears essential in each.} But there is at least a well-established formalism to study when the quantum state can \emph{safely be treated} as probabilistic: the \emph{decoherent-histories formalism} (Griffiths~\citeyear{griffiths,Griffiths1993}, Gell-Mann and Hartle~\citeyear{gellmannhartle,gellmannhartle93}, \citeN{halliwellreview}, \citeN{omnes92}) which I will briefly describe for readers unfamiliar with it.

The ingredients for the decoherent-histories formalism (in a slightly simplified form sufficient for our purposes) are:
\begin{enumerate}
\item A quantum theory, defined on some Hilbert space \mc{H} and with dynamics given by a family of unitary operators $\op{U}(t)$;
\item An increasing sequence of times $t_0,t_1, \ldots t_n$;
\item An orthogonal family of projectors $\{\op{\Pi}_J\}$, such that $\sum_J\op{\Pi}_J=\id$, representing mutually-exclusive properties of the system;
\item An initial state --- that is, a density operator $\denop$ on \mc{H} --- representing the system's state at time $t_0$.
\end{enumerate}
A \emph{history} $h$ of the system is then a function from times $t_i$ in the sequence to operators $h(t_i)$, where each $h(t_i)$ is a sum of some subset of the projectors 
$\op{\Pi}_J$, and hence represents a possibly-coarse-grained property. The intended interpretation is that $h(t_i)$ is the property possessed by the system at time $t_i$. The histories are then exactly the  sequences of (possibly-coarse-grained) ideal measurements (that is, measurements of the kind we discussed above) of the $\op{\Pi}_J$. To each history corresponds a \emph{history operator}
\be
\op{H}(h)=h(t_n) \op{U}(t_n-t_{n-1})h(t_{n-1}) \cdots \op{U}(t_2-t_1) h(t_1) \op{U}(t_1-t_0) h(t_0).
\ee
We now define the \emph{decoherence functional} as
\be
D_\rho(h,h') = \tr (\rho\opad{H}(h)\op{H}(h')).
\ee
(Note that the functional depends on the initial state as well as on the histories.)

The decoherence functional has two key properties. Firstly, $D_\rho(h,h)\equiv \Pr_\rho(h)$ is the Born-rule probability that a sequence of ideal measurements of the properties represented by the $\op{\Pi}_J$ will give a sequence of outcomes of which history $h$ is  a coarse-graining. Secondly, suppose that $h_1$ and $h_2$ satisfy $h_1(t)=h_2(t)$ except for $t=t_i$, and that $h_1(t_i)$ and $h_2(t_i)$ are orthogonal. Then we can define a history $h_1+h_2$ by
\[
h(t)=h_1(t)=h_2(t) \,\,\,\,\, t \neq t_i
\]
\be
h(t_i)=h_1(t_i)+h_2(t_i).
\ee
Classically, we would expect $\Pr_\rho(h_1+h_2)=\Pr_\rho(h_1)+\Pr_\rho(h_2)$, but quantum-mechanically we find
\begin{eqnarray}
\nonumber
\Pr_\rho(h_1+h_2) &=& \Pr_\rho(h_1)+\Pr_\rho(h_2) + D_\rho(h_1,h_2)+D_\rho(h_2,h_1) \\
&= &\Pr_\rho(h_1)+\Pr_\rho(h_2) + 2\real{D_\rho(h_1,h_2)}.
\end{eqnarray}
So the real part of the decoherence functional between \emph{different} histories is a measure of the interference between those histories. If $\real{D_\rho(h,h')}=0$ for distinct histories $h,h'$, we say that the space of histories is \emph{consistent} --- meaning that we can consistently use the decoherence function to assign probabilities to those histories (at least formally). In practice, the weaker criterion of \emph{approximate consistency} --- $\real D_\rho(h,h')\simeq 0$ for $h\neq h'$ --- is generally taken to be sufficient, guaranteeing as it does that $D(h,h)$ can for all practical purposes be treated as a probability. A more calculationally useful, though strictly speaking overly strong, variation on this is \emph{medium decoherence}, the requirement that $ D_\rho(h,h')\simeq 0$.

Of course, identifying formal conditions under which the quantum state can be treated as probabilistic does not in itself resolve the deep conceptual puzzle about how to understand a description of a system that is probabilistic in some contexts and seems to represent the objective physical state of the system in other contexts. (My own view, defended at length in \citeN{wallacebook}, is that we need to understand the quantum state as representating the objective, physical state of the system in all contexts, and to see the medium decoherence condition as a necessary condition to interpret that state as realising a branching, emergently probabilistic structure along the lines of Everett's \citeyear{everett} many-worlds theory --- but I shall not presuppose that view here.) But however this conceptual problem is resolved, to make contact with the empirical results of quantum mechanics it will have to --- somehow --- legitimate the normal practice of interpreting the quantum state as probabilistic in the absence of interference. So I will assume, for the purposes of this article, that we can interpret decoherent quantum histories probabilistically.

So much for the \emph{definition} of decoherence and consistency. The most commonly discussed \emph{dynamical origin} of consistency is \emph{environment-induced decoherence} (Zeh~\citeyear{zeh1973,zeh93}, \citeN{jooszeh85}, Zurek~\citeyear{zurek91,zurekroughguide}; for reviews, see Joos \emph{et al}~\citeyear{joosetal} or \citeN{schlosshauerbook}), where the family of projectors tracks only properties of some system and traces over the properties of an environment (this was proposed as a mechanism for SSB by \citeN{simonius}). 

But (as stressed by, \egc, Halliwell~\citeyear{halliwellhydrodynamic,halliwellconference}) another source of decoherence is \emph{conservation}: if some quantity is approximately conserved by the dynamics, the histories of that quantity will decohere. 

In more detail: suppose that we choose our family of projectors $\{\op{\Pi}_J\}$ to satisfy $\op{U}(t)\op{\Pi}_J\opad{U}(t)\simeq \op{\Pi}_J$. Then we can distinguish between constant histories (for which $h(t)=h(t')$) and non-constant histories. It is easy to check that the decoherence functional now simplifies dramatically, to
\begin{eqnarray}
\nonumber
D_\rho(h,h') & \simeq & 0  \,\,\,\, (h\mbox{  or  }\emph{h'}\mbox{   non-constant or }h\neq h')\\
&\simeq& \tr(\denop h(t_0)) \,\,\,\,\mbox{   (constant }h,\,h=h')
\end{eqnarray}
In this situation, we can treat the system as decomposing into dynamically isolated sectors, each characterised by a particular value of the quantity represented by the $\{\op{\Pi}_i\}$, and each occurring with probability $\tr(\denop \op{\Pi}_i)$. As we will shortly see, this conservation-based conception of decoherence  plays a central role in SSB.

\section{Decoherence and spontaneous symmetry breaking}

Let's consider the single- and double-well potentials from a decoherent-histories viewpoint. We choose as our family of projectors the two projectors $\op{\Pi}_L$, $\op{\Pi}_R$ that project onto states localised on (respectively) the left and right sides of the origin, choose as times the equally-spaced times $0,\tau, 2\tau, \ldots n\tau$, and continue to work in the approximation introduced in Section \ref{singleparticle} where we consider only the ground and first excited states $\ket{\psi_0},\ket{\psi_1}$ (with energy separation $\hbar \omega$), so that we can approximate
\[
\op{\Pi}_L = \frac{1}{2}(\ket{\psi_0}-\ket{\psi_1})(\bra{\psi_0}-\bra{\psi_1})
\]
\[
\op{\Pi}_R = \frac{1}{2}(\ket{\psi_0}+\ket{\psi_1})(\bra{\psi_0}+\bra{\psi_1}).
\]
Given the results of Section \ref{singleparticle}, if $t \ll 1/\omega$ then $\op{U}(t)\op{\Pi}_{(L/R)}\opad{U}(t)\simeq \op{\Pi}_{(L/R)}$, so that the conservation conditions for consistency are satisfied. We can then reinterpret section \ref{singleparticle} as telling us that for the double-well potential, unless we study it on timescales long compared to the between-well tunnelling timescale, the left-side and right-side histories decohere and the system can thus be considered as a probabilistic mixture of a particle localised on the left side and one localised on the right side. Furthermore, this should continue to hold if we replace the ground state with a thermal state with expected energy much lower than the height of the energy barrier between the two wells (since such a state is a thermal mixture of energy eigenstates that come in symmetric/antisymmetric pairs with extremely small energy gaps between them, just as for the ground state).

It is important to note that the system is by no means \emph{completely} decoherent, nor is its full dynamics even approximately classical. The projectors $\op{\Pi}_L$, $\op{\Pi}_R$ are extremely coarse-grained, and the left-side and right-side sectors can still display complicated quantum-mechanical dynamics (the dynamics of a perturbed harmonic oscillator, in particular). The decoherence between left-side and right-side histories allows the quantum state to be treated as a probabilistic mixture of left-side and right-side quantum states, but those states are not further decohered, at least by any process internal to the oscillator.

As in the classical case, things become conceptually clearer (if technically more complicated) if we move to the case of a field theory with a double-well potential. The ground state of that system is a classic ``Schr\"{o}dinger-cat'' state: a superposition of two wavefunctions localised around macroscopically distinct field configurations. As such, even before considering its dynamics we can see that the system displays the first two (closely related) characteristics of SSB identified in Section  \ref{classicalcase}: the field values of different spatial points will be closely correlated, and the spatially averaged field strength, though it has expected value zero, will with probability $\sim 1$ be found to be macroscopically different from zero. 

As for the dynamical behavior of the system: formally speaking, the analysis we have already made applies here too, simply substituting for the energy gap $\Delta E$ the result (\ref{energygap-field}) obtained in section \ref{groundstate}. Provided that we are considering the dynamics on timescales $t \ll \hbar/\Delta E$, and provided the system's initial state has expected energy well below the energy required for thermal transitions across the potential barrier, the system will decohere into two sectors characterised by positive and negative average field values. Each sector will look like a scalar quantum field theory describing excitations around the respective minima, and dynamical coupling between the two sectors will be so negligible that interference between them may be neglected, justifying the probabilistic interpretation. The two sectors individually, however, will be fully quantum-mechanical (at least in the absence of some further decoherence mechanism).

It is instructive to quantify the timescales on which this treatment is appropriate, for a solid-state system. They are controlled by the instanton action $S$ defined by equation (\ref{instanton-field}), according to which the action is proportional to the volume of the system. We can obtain an extremely crude estimate simply via dimensional analysis: solid-state systems are characterised by typical masses $m$ around the proton mass $m_p$, typical energies $E$ around the $1$eV typical of atomic systems, and typical lengthscales $l$ around the $10^{-10}$ m interatomic distance. The most general way to combine these to get a quantity of dimension (action/volume) is
\be
S/V \alpha \frac{\sqrt{mE}}{l^2} \sim \alpha  \times 10^{-3} \mathrm{Js/m^3}
\ee
in SI units, where $\alpha$ is a dimensionless constant that (in the absence of really remarkable emergent effects) will be within a few orders of magnitude of unity. (It will be apparent in a moment why this level of crudeness is acceptable). The overall dependence of $\Delta E$ on $S/V$ is, to leading order and ignoring dimensionless factors,
\be
\Delta E \sim E \times \e{-SL^3/\hbar V}.
\ee
For (say) a one-cubic-millimeter sample of a solid, $SL^3/\hbar V\sim 10^{22}$, so that 
\be
\Delta E \sim \e{-10^{22}}.
\ee
We would then expect the two symmetry-breaking histories to be effectively decoherent for timescales short compared to $\e{10^{22}}\mathrm{s}$. To call this length of time ``large compared to the age of the Universe'' would be a ridiculous understatement. So even for a finite --- indeed, quite small --- sample, and even in the complete absence of external perturbations, we can to a ridiculously good approximation treat a system below the phase-transition energy as being in a probabilistic mixture of two symmetry-breaking states.

The decoherent-histories framework can also be applied to cases of spontaneously-broken \emph{continuous} symmetry, although there is no some arbitrariness in the definition of the history projectors. Starting from the nonlinear-sigma-model approximation in equation (\ref{nonlinear-sigma}), we need to define projectors with respect to the spatially-averaged phase $\bar\theta$, and we can do so by dividing the unit circle into equal sections of angular width $\Delta$ and defining projectors onto states with spatially-averaged phase having support in those sections. In this case (see Appendix for details), the conservation condition for consistent histories is satisfied on timescales $t$ satisfying
\be
t \ll  \frac{\Delta^2 L^3 \rho_0^2}{\hbar}.
\ee
This goes to infinity as $L\rightarrow \infty$; indeed, we can take a simultaneous limit of $\Delta\rightarrow 0$ and $L \rightarrow \infty$, and recover the fact that in the infinite limit the dynamically-isolated sectors are identified with a sharp value of the spatially-averaged phase. Furthermore, although it does not go to infinity nearly so quickly as in the discrete case (it diverges linearly, rather than exponentially, with system volume) there are good reasons to think that it is sufficiently quick to explain the actual phenomenology of finite systems.  Using the same crude dimensional-analysis approach used previously, we find that $\rho_0^2$ has order of magnitude
\be
\rho_0^2 \sim \frac{\hbar}{l}\left(\frac{m}{E}\right)^{1/2}
\ee
so that the timescale of validity for the consistent-history approach is
\be
t_{CH} \sim l \left(\frac{m}{E}\right)^{1/2} \left(\frac{L}{l_0}\right)^3 \Delta^2 \sim 10^{16}L^3 \Delta^2
\ee
in SI units. As a rough estimate for the timescales $t_\mu$ relevant for microphysical processes in a solid, we can take $t_\mu=L/c$ where $c\sim 10^3 \mathrm{m/s}$ is the speed of sound in the solid. So we can treat the histories as decohering when studying solids provided that $t_\mu\ll t_{CH}$, \ie
\be
\Delta^2 \gg 10^{-19}/L^2.
\ee
For the one-cubic-millimeter sample we previously considered, this becomes
\be
\Delta^2 \gg 10^{-11}
\ee
as a condition for applying the decoherent-histories formalism. 
SSB will occur provided $\Delta \ll 2\pi$, so (on this admittedly crude estimate) we can treat the sample as spontaneously breaking symmetry on timescales far longer than the relevant microphysical timescales for the system, even for a small sample.

For a sufficiently hot sample, and or a sufficiently energetic external environment, we would also expect environment-induced decoherence to induce SSB, because either thermally-excited Goldstone bosons or the external environment will record the spatially-averaged phase. In principle, an extremely small sample (only a few dozen atoms across) cooled to a sufficiently low temperature and sufficiently isolated from its environment might not be well treated by assuming broken symmetry, but systems like this in any case lie outside the class of systems that we would expect to be well treated by conventional field-theoretic methods.

\section{Conclusion}

According to Butterfield's account, we can only understand SSB via the infinite limit if we can identify some feature of finite systems which corresponds at least approximately to SSB, with the approximation becoming exact in the limit. Degeneracy of the ground state is not such a feature, appearing as it does only for actually-infinite systems. Explicit symmetry breaking by external perturbations has the right formal structure, but seems incompatible with features of the actual phenomenology of symmetry breaking in finite systems.

The real signature of SSB in finite systems is the breakdown of those systems' state spaces, at least at low energies, into sectors that are approximately dynamically isolated, as identified using the tools of the decoherent-histories approach. The approximation becomes exact in the infinite limit: we can interpret the inequivalent representations that correspond to spontaneous symmetry breaking as infinite-volume limits of the approximately-isolated decoherent sectors. (In that limit, different-temperature regimes also correspond to inequivalent representations, so the restriction to low energy ceases to apply: the symmetry is still restored at high temperature, but this is seen at the level of representations rather than within any given representation.)

From this perspective, the common identification of SSB with ground-state degeneracy looks misleading. Ground state degeneracy matters because, in an infinite system, it is roughly-speaking equivalent to the existence of inequivalent, symmetry-related vacuum representations, but it is the existence of those representations, not ground-state degeneracy per se, which gives rise to the phenomenology generally associated to spontaneous symmetry breaking.

Also from this perspective, we can understand the close relation between SSB and the quantum measurement problem, even without appeal to the infinite limit. A quantum system, initially in a perfectly symmetric quantum state describing thermal equilibrium above the critical temperature, remains in that perfectly symmetric state as it is cooled to absolute zero. What changes is that it realises that symmetry, at sufficiently low temperatures, by being a superposition of macroscopically distinct states, each individually non-symmetric, and each acting as the \emph{de facto} ground state of a space of states dynamically isolated from those spaces built on other terms in the superposition. Whatever our preferred resolution to the measurement problem, it will need to permit us to treat such macroscopic superpositions as probabilistic mixtures; whether that is achieved through some new dynamics or new variables, through some probablistic reinterpretation of the quantum state, or through the emergent branching of the Everett interpretation, it will also provide us with what we need to resolve the related puzzle of spontaneous symmetry breaking in finite systems.

\section*{Acknowledgements}

I'm extremely grateful to Jeremy Butterfield for a close and careful reading of the manuscript, which has improved it substantially.

\section*{Appendix}

Consider a quantum-mechanical particle in one dimension with Hamiltonian
\be
\op{H}=\op{P}^2/2m.
\ee
and initially in a square-wave state,
\begin{eqnarray}
\nonumber
\bk{x}{\psi}&=\frac{1}{\sqrt{\Delta}}&(|x|<\Delta/2)\\
&= 0 &(|x>\Delta/2)
\end{eqnarray}
In a momentum basis, this state is
\be
\bk{k}{\psi}=\sqrt{\frac{2}{\pi \Delta}}\frac{\sin(k\Delta/2)}{k}.
\ee
So
\be
\matel{\psi}{\exp(-it\op{H}/\hbar)}{\psi}= \frac{2}{\pi \Delta}\int \dr{k}\e{-itk^2/2m\hbar}\frac{\sin^2(k\Delta/2)}{k^2}
\ee
or, after a change of variable,
\be
\matel{\psi}{\exp(-it\op{H}/\hbar)}{\psi} = \frac{2}{\pi}\int \dr{z}\e{-itz^2/2m\Delta^2\hbar}\frac{\sin^2(z)}{z^2}.
\ee
If $t\ll m\Delta^2\hbar$, then the exponent in the integral will vary only slowly in the region where $\sin^2(z)/z^2$ is large, and so may be approximated as constant. In this regime, the integral is $t$-independent and so can be evaluated at $t=0$, where it is trivially equal to 1. So a square wave-packet does not spread out appreciably provided that this condition is satisfied.

The spatially-averaged phase of the field theory describing spontaneous symmetry breaking can be approximated by this system if we put $m=\rho_0^2 L^3$, provided we take $\Delta\ll 2\pi$ (so that the system may be approximated as permitting phase values in $(-\infty,\infty)$ and not just $[0,2\pi)$. We then have our condition for consistency of histories defined by projections of width $\Delta$:
\be
t \ll \frac{\Delta^2 L^3 \rho_0^2}{\hbar}.
\ee


\end{document}